\documentclass[aps,prl,twocolumn,showpacs,floatfix,10pt]{revtex4-1}
\usepackage{amsmath}    % need for subequations
\usepackage{graphicx}
\usepackage{hyperref}
\usepackage[english]{babel} 

\begin{document}

\title{Strong coupling of the cyclotron motion of surface electrons on liquid helium to a microwave cavity}

\author{L.V. Abdurakhimov}
\email{la396@cam.ac.uk}
\altaffiliation[Current address: ]{Microelectronics Group, Cavendish Laboratory, Cambridge University, Cambridge CB3 0HE, United Kingdom}
\author{R. Yamashiro}
\author{A.O. Badrutdinov}
\author{D. Konstantinov}
\affiliation{Okinawa Institute of Science and Technology (OIST) Graduate University, Onna, Okinawa 904-0495, Japan}

\date{\today}

\begin{abstract}
The strong coupling regime is observed in a system of two-dimensional electrons whose cyclotron motion is coupled to an electro-magnetic mode in a Fabry-Perot cavity resonator. The Rabi splitting of eigenfrequencies of coupled motion is observed both in the cavity reflection spectrum and ac current of electrons, the latter probed by measuring their bolometric photoresponse. Despite the fact that similar observations of the Rabi splitting in many-particle systems have been described as a quantum-mechanical effect, we show that the observed splitting can be explained completely by a model based on the classical electrodynamics. 
\end{abstract}


\maketitle

\paragraph{Introduction.} Recent years were marked by a significant interest in the strong coupling of a collection of quantum particles to the electro-magnetic modes of a resonator. Besides traditional systems used in cavity QED experiments such as Rydberg atoms~\cite{Kaluzny1983,Zhu1990,Thompson1992}, strong coupling regime has been recently studied in various paramagnetic~\cite{Chiorescu2010,Schuster2010,Kubo2010,Amsuss2011,Abe2011} and ferromagnetic~\cite{Huebl2013,Tabuchi2014,Zhang2014} electron spin ensembles, a coupled nuclear-electron spin system~\cite{Abdurakhimov2015}, as well as two-dimensional electron systems (2DES) in semiconductors~\cite{Hagenmuller2010, Scalari2012, Muravev2011, *Muravev2013, QiZhang2016} and graphene~\cite{Hagenmuller2012,Chirolli2012}. The hallmark of the strong coupling regime is the splitting in the resonator spectrum revealed in the signal reflected from or transmitted through the resonator. In case of a collection of $N$ quantum particles this splitting scales as $\sqrt{N}$ and is referred to as the Rabi splitting~\cite{Agarwal1984,Chiorescu2010}.

Besides general interest in the fundamental problem of light-matter interaction, the particular interest in the strong coupling regime comes from the quantum information processing as strong coupling to a high-Q resonator enables coherent information transfer between, for example, a qubit and spin system excitations~\cite{Imamoglu2009}. Therefore, most of the recent observations of strong coupling have been interpreted as pure quantum phenomena. However, it is rarely mentioned that the strong coupling between large N-particle ensemble and the coherent state of electromagnetic mode in a resonator can be described completely classically in many cases~\cite{Zhu1990,Muravev2011,*Muravev2013,Bai2015}. Indeed, one needs to introduce non-linearity to a strongly-coupled quantum system in order to create pure quantum states (e.g., a superconducting qubit can be used for creation of non-classical states~\cite{Hofheinz2008}). Otherwise, in a linear system, such as coupled system of N-particle ensembles and electromagnetic cavity mode, the problem is equivalent to two coupled harmonic oscillators which exhibit the normal mode splitting when the eigen frequencies of uncoupled oscillators coincide~\cite{HarocheRaimond}.

In this Letter, we report observation of the strong coupling between cyclotron mode of 2D electrons on the surface of liquid helium and a 3D microwave cavity resonator. The splitting in the eigen spectrum of coupled motion is observed in the cavity reflection signal, as well as in the ac current of electrons detected by measuring their bolometric photoresponse. A simple model that uses, on the one hand, an expression for the ac conductivity  of electrons and, on the other hand, the classical equations for electro-magnetic field in the cavity accounts for all experimental features including the observed splitting. The square-root scaling of the splitting with the number of electrons follows naturally from our model.  Thus, our work reproduces all features of the strong coupling regime for a large N-particle 2DES but puts it on a completely classical ground.

\begin{figure}
\includegraphics{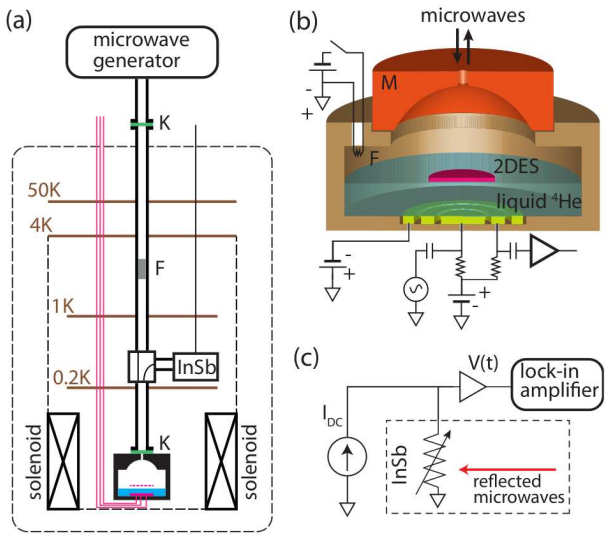}
\caption{\label{fig1} (color online) (a) Schematic diagram of the experimental setup. (b) Sketch of the experimental cell and the Fabry-Perot resonator. (c) Circuit for microwave power measurements.}
\end{figure}

\paragraph{Experimental setup.} Measurements were performed at temperature of $T\approx 0.2$\,K in a dilution refrigerator (Fig.~\ref{fig1}a). Liquid helium-4 was condensed into a vacuum-tight cylindrical copper cell with internal diameter of 40\,mm (Fig.~\ref{fig1}b). The cell contained a semiconfocal Fabry-Perot resonator formed by a top spherical mirror and a flat gold-film mirror at the cell bottom. The distance between two mirrors was $D=7.4$\,mm. The flat mirror was made in a form of three concentric electrodes (the Corbino disk). Further details can be found elsewhere~\cite{Yamashiro2015}. The Fabry-Perot resonator was operated at TEM$_{003}$ mode which can be described by the Gaussian beam distribution~\cite{Kogelnik1966, TEM_notation}. The beam waist $w_0$ was calculated to be about 2\,mm. The microwave electric field was parallel to the liquid helium surface, and the liquid surface was located in the antinode of TEM$_{003}$ mode (at distance $h\approx 0.85$~mm above the flat mirror) at which the amplitude of the microwave electric field was maximal. The cavity resonance frequency was $\omega_\textrm{r}/ 2 \pi \approx 88.4$\,GHz, and Q-factor was measured to be about $900$ at low temperatures. The resonator was probed by pulse modulated microwave pumping at frequency $\omega$ with modulation rate of 5\,kHz. The microwave power $P_\textrm{r}$ reflected from the resonator was detected by an indium antimonide (InSb) hot-electron bolometer and synchronously demodulated by a lock-in amplifier (Fig.~\ref{fig1}c).  Sensitivity of the InSb bolometer was measured to be about 700~V/W.

Free electrons were injected into the cell via thermal emission from a tungsten filament mounted inside the cell (see Fig.~\ref{fig1}b) while a positive voltage $V_\textrm{B}$ was applied to the center and middle electrodes of the Corbino disk. In addition to the electrostatic potential created by biased electrodes, electrons experience a long-range attractive force towards free surface of liquid helium  due to its polarizability. On the other hand, electrons are affected by a short-range repulsive barrier at the liquid surface due to the negative affinity of an electron to a $^4$He atom caused by the Pauli exclusion principle. Due to the resultant potential well seen by electrons, a 2DES is formed on the free surface of liquid helium~\cite{Andrei,MonarkhaKono}. These surface electrons (SEs) can freely move along the helium surface, but their vertical motion is quantized. The corresponding surface energy levels are described by the hydrogen-like spectrum. The energy difference between the ground surface level and the first excited level is about 0.55\,meV ($\approx 6$\,K in terms of temperature)  in zero electric pressing field ($V_\textrm{B}=0$), and it increases with the increase of pressing field due to the linear Stark effect. Therefore, for typical temperatures used in the experiment SEs occupy the ground surface level. Density of electrons is determined from the condition of complete screening of electric field above the surface, $n_e = \epsilon_0 \epsilon V_B/ e h$, where $\epsilon_0$ is the vacuum permittivity, $\epsilon =1.057$ is liquid helium dielectric constant, and $e$ is the electron charge. The negatively biased outer bottom electrode was used as a guard ring to prevent electrons from escape. The magnetic field $B$ perpendicular to the liquid surface was created by a superconducting solenoid, and cyclotron resonance could be excited by the microwave electrical field when the frequency $\omega$ matched the cyclotron frequency $\omega_c=eB/m_e$, where $m_e$ is the free electron mass.
\begin{figure*}
\begin{center}
\includegraphics[width=14cm,keepaspectratio]{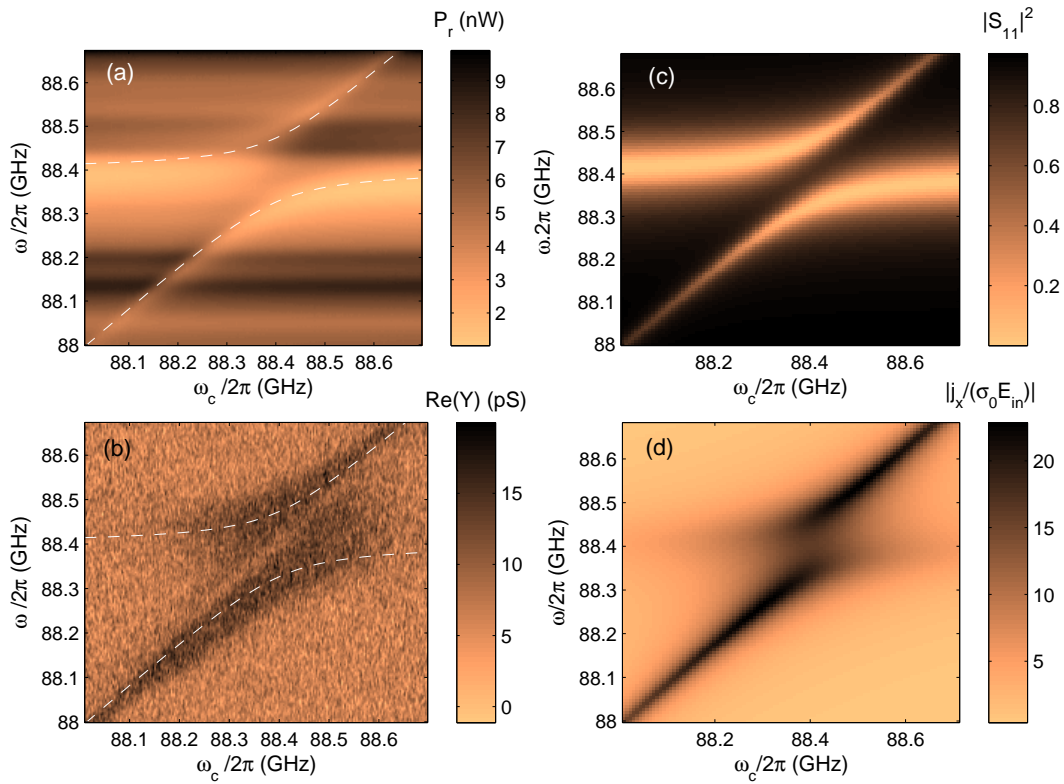}
\end{center}
\caption{\label{fig2} (color online) Left side: Results of simultaneous measurements of (a) microwave power reflected from the cavity, $P_r$, and (b) real part of dc admittance of the cell, $\textrm{Re}(Y)$, as a function of the microwave probe frequency $\omega$ and magnetic field in units of $\omega_c$. Electron density is $n_e \approx 2 \times 10^8$\,cm$^{-2}$, and microwave probe power is $P_{in} \approx 180$~nW. Dashed lines correspond to calculated eigenfrequencies of cavity-field-2DES coupled motion with coupling strength $g/2\pi \approx 77$~MHz. The horizontal stripes on Fig.~\ref{fig2}a are caused by a parasitic standing wave formed in the waveguide due to slight impedance mismatching. Right side: Absolute value of (c) normalized reflected power $|S_{11}|^2$ and (d) normalized electron current density $j_x/(\sigma_0 E_{in})$ calculated using model described in the text for $n_e = 10^8$\,cm$^{-2}$, $\gamma=4\times 10^8$~s$^{-1}$, and $\nu=10^8$~s$^{-1}$.}
\end{figure*}
\paragraph{Strong coupling.} Figure~\ref{fig2}a shows the reflected microwave power $P_\textrm{r}$ obtained with SEs in the cavity under typical experimental conditions and plotted as a function of the microwave probe frequency $\omega$ and external magnetic field in units of  $\omega_c$. A pronounced avoided crossing in the cavity resonance is found near the degeneracy point, that is the point where the uncoupled cavity mode $\omega_r$ would otherwise cross the uncoupled cyclotron-resonance mode $\omega_c$. Thus, the observed anticrossing behavior represents normal-mode splitting in the coupled system of two oscillators: the cavity mode and the cyclotron motion of SEs. For the data presented in Fig.~\ref{fig2}a, we find that the value of the splitting between two normal modes at the degeneracy point is about $2g/2\pi = 154$ ~MHz. The full linewidth of the cavity mode is about $\gamma/2\pi = 100$~MHz, and the full linewidth of the cyclotron mode is approximately $\nu/2\pi = 20$~MHz. Thus, the cooperativity $C = 4 g^2 / \gamma \nu \approx 12$ is larger than unity, and, hence, strong coupling regime is realized in our experiment.

\begin{figure*}
\begin{center}
\includegraphics[width=14cm,keepaspectratio]{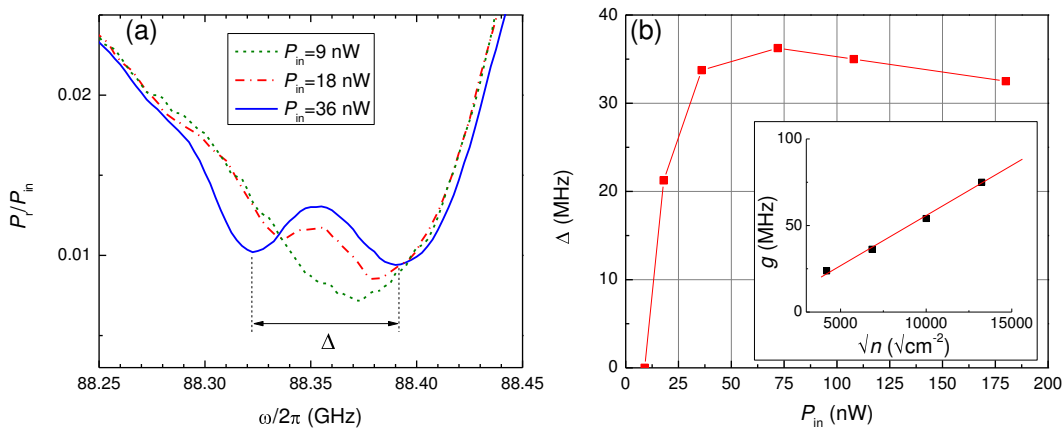}
\end{center}
\caption{\label{fig3} (color online) (a) Normalized spectra of microwave power reflected from the cavity at several values of input power. Measurements were performed at magnetic field corresponding to pure cyclotron resonance frequency $\omega_c=88.35$~GHz. Electron density is $n_e \approx 5\times 10^7$~cm$^{-2}$. The splitting appears only at high powers due to power-induced narrowing of cyclotron resonance. (b) Power dependence of coupling strength $g$ measured as a distance $\Delta$ between two minima of $P_\textrm{r}$ near the degeneracy point. Inset: Dependence of coupling strength $g$ on square root of electron density, $\sqrt{n_e}$, at input power of $P_{in} \approx 70$~nW.}
\end{figure*}

The observed avoided crossing is consistent with the Rabi splitting effect, which is typically discussed in the context of similar experiments on strong coupling between quantum particle ensembles and cavity modes~\cite{Chiorescu2010,Schuster2010,Kubo2010,Amsuss2011,Abe2011,Huebl2013,Tabuchi2014,Zhang2014,Abdurakhimov2015,Hagenmuller2010,Scalari2012}. In our experiment, the latter would be given by the coupling constant $g$ in the form $g = g_0 \sqrt{N_e}$, where $g_0$ is the coupling strength for a single electron, and $N_e$ is the total number of electrons coupled to the cavity mode. For nondegenerate SEs occupying the lowest energy level of cyclotron motion the coupling strength is given by $g = (e l_\textrm{B} E_0)/\hbar$, where $l_\textrm{B} = \sqrt{\hbar / m_e \omega_c}$ is the magnetic length and $E_0$ is the vacuum RMS electric field in the cavity. The latter can be  estimated as $E_0= \sqrt{\hbar \omega/2\epsilon_0 V}$, where $V$ is the cavity mode volume. For comparison with the experiment, the total number $N_e$ can be roughly estimated as $N_e = n_e \times S$ where $S$ is the characteristic spot size of microwave Gaussian beam at the liquid helium surface, $S =  \pi w_0^2 \approx 12.6$\,mm$^2$. The mode volume $V_m$ for Gaussian beam, described by distribution $E(r,z)=E_0 \times f(r,z)$, was estimated numerically by integration over the cavity volume as $V_m = \int f^2(r,z) 2 \pi r dr dz \approx 0.02$\,cm$^3$. Thus, we obtain estimation value $g/2 \pi \approx 200$\,MHz, which is comparable with the experimental value $g/2 \pi = 77$\,MHz. The observed dependence of the splitting on the electron density is consistent with $\sqrt{N_e}$ scaling (see insert in Fig.~\ref{fig3}b). 

In addition to microwave cavity reflection measurements, we also performed simultaneous detection of the ac current in the 2DES induced by cavity field. Because it is very difficult to measure such a high frequency current directly, we employed its detection using electron bolometric photoresponse~\cite{Edelman1977}. The method is based on the effect of heating induced by ac current of SEs on the electron dc resistivity~\cite{Aoki1980}. The latter could be probed by measuring the complex admittance of the cell by the capacitive-coupling (Sommer-Tanner) technique ~\cite{Mehrotra1987,Wilen1988} using the central and middle electrodes of the Corbino disk, see Fig.~\ref{fig1}b. The real part of admittance, which is proportional to electron dc resistivity, is plotted in Fig.~\ref{fig2}b as a function of $\omega$ and $\omega_c$. The splitting of cyclotron resonance mode $\omega=\omega_c$, as well as its significant broadening, is clearly observed near the degeneracy point. This measurement directly confirms that the coupling to microwave cavity mode modifies the electron cyclotron motion and introduces additional damping due to decay of the cavity field.

\paragraph{Theoretical model.} To understand the above results, we developed a model of 2DES in a simplified cavity resonator, basing on the approach suggested in~\cite{Shikin2002}. In our model, the motion of coupled system is described by the following equations~\cite{supplemental}
\begin{align}
\begin{pmatrix}
 \frac{D}{2c}\left(\omega - \omega_r + i\gamma\right)  &   i\eta_0/2 \\
-\frac{n_e e^2}{m_e} (\nu-i\omega) &  \quad (\omega+i\nu)^2 - \omega_c^2 
\end{pmatrix}
\begin{pmatrix}
E_x \\
j_x 
\end{pmatrix}
=
\begin{pmatrix}
E_{\textrm{in}} \\
0 
\end{pmatrix},
\label{eq:coupled}
\end{align}
where $\gamma$ is the total loss rate of the resonator, $D$ is the distance between resonator mirrors, $c$ is the speed of light, $\eta_0$ is the intrinsic impedance of vacuum, $\nu$ is the cyclotron resonance linewidth determined by electron scattering and Coulomb interaction between electrons~\cite{MonarkhaKono,Dykman1997,Teske1999,Monarkha2000}, and $E_\textrm{in}\propto \sqrt{P_\textrm{in}}$ is the electric field component of microwave probe pump. The above equations relate the microwave electric field $E_x$ and electron current density $j_x$ in the 2DES plane through classical electrodynamics relations and electron ac conductivity $\sigma_{xx}=j_x/E_x$. For $E_\textrm{in}$=0, the solutions of Eq.~(\ref{eq:coupled}) are two damped eigenmodes with frequencies plotted in Figs.~\ref{fig2}a,b by dashed lines. 

For $E_\textrm{in}\neq 0$, Eq.~(\ref{eq:coupled}) can be readily solved for given pump frequency $\omega$. The $S_{11}$-parameter is given by~\cite{supplemental}
\begin{equation}
S_{11} = 1+\frac{2(\gamma_{ext}-i\delta\omega_{ext})}{i(\omega - \omega_r)^2 - (\gamma_{int} + \gamma_{ext}) - \frac{\sigma_{xx}}{\varepsilon_0 D}},
\label{s11}
\end{equation}
where $\gamma_\textrm{int}$ and $\gamma_\textrm{ext}$ are the internal and external loss rates of the resonator, $\gamma=\gamma_\textrm{int} + \gamma_\textrm{ext}$, and $\delta\omega_{ext}$ is the resonator frequency shift due to external coupling. For the sake of comparison with the experiment, $|S_{11}|^2$ and normalized current density $|j_x/(\sigma_0 E_\textrm{in})|$, where $\sigma_0=n_e e^2/(m_e\nu)$, are shown in Figs.~\ref{fig2}c,d. Clearly, our model completely accounts for all experimental observations.

According to our model the normal mode splitting at the degeneracy point is given by $2g=2\sqrt{\frac{n_e e^2}{2\varepsilon_0 m_eD}}$~\cite{supplemental}. Note that this result coincides with the expression for the Rabi splitting in terms of the vacuum RMS electric field $E_0$. Indeed, after multiplying and dividing the above result by $\hbar$, it is straightforward to rewrite it in the form $2g=(2el_B E_0)/\hbar$  (see~\cite{supplemental}). We can also represent the coupling constant $g$ in the form $g\propto \sqrt{\alpha n_e}$, where $\alpha=e^2/c\hbar$ is the fine structure constant. A similar ``quantum'' representation for the coupling constant (denoted there by $\Omega$) was used in Ref.~\cite{Scalari2012} to describe strong coupling between the cyclotron transition of 2DES and THz resonators. Thus, we demonstrate that, similar to our work, results of Ref.~\cite{Scalari2012} can be explained by a classical model as well. 

A peculiar feature of our experimental results is a strong dependence of the coupling regime on the microwave probe power $P_\textrm{in}$. The normal-mode splitting becomes noticeable only at sufficiently high power, as shown on Figure~\ref{fig3}a,b. We suppose that the observed behavior is related to the strong non-monotonic dependence of cyclotron resonance linewidth $\nu$ on the microwave power. Indeed, the so called Coulomb narrowing of the cyclotron resonance linewidth with microwave power was observed and explained in terms of electron heating and increase of many-electron fluctuating electric field experienced by an electron from its neighbours~\cite{Teske1999}. This field leads to suppression of electron scattering within the Landau level (LL), and, therefore, a reduction in $\nu$. For small powers, the cyclotron resonance linewidth is about 100~MHz~\cite{Edelman1977} and, since $g<\nu,\,\gamma$, the system is not in the strong coupling regime and the normal mode splitting is not observed. With increasing power, reduction in $\nu$ allows to reach the strong coupling regime, and the normal-mode splitting is observed. Further increase in power leads to further increase of many-electron fluctuating electric field, which eventually assists the inter-LLs scattering, resulting in an increase in $\nu$~\cite{Teske1999,Monarkha2000}. Correspondingly, the observed normal-mode splitting diminishes, as shown in Fig.~\ref{fig3}b. The dependence of observed splitting on $\nu$ obtained from our model is in good agreement with the experimental observation~\cite{supplemental}.

\paragraph{Conclusions.} We report observation of the strong coupling regime between a collection of 2D electrons on liquid helium and microwave cavity mode. The reported normal-mode splitting, also referred to as the Rabi splitting, is observed in both microwave response and electron transport measurements and shows correct scaling with the number of particles. We demonstrate that, in contrast with usual quantum-mechanical description of similar observations in other experiments, our result can be completely explained by a classical model. Similar classical treatment should be able to account for observations of the strong coupling in other linear systems. We note that adding a nonlinear quantum system, such as a qubit, to our experiment can provide possibility to use the pure bosonic system of quantum oscillators on liquid helium for cavity QED experiments and quantum information processing. In addition, the presented experimental method provides possibility to study intriguing radiation-induced magneto-transport phenomena such as zero-conductance~\cite{Konstantinov2010} and incompressible states~\cite{Chepelianskii2015} of 2DES on liquid helium in the regime of strong coupling to radiation field. 

\begin{acknowledgments}
The work was supported by an internal grant from Okinawa Institute of Science and Technology (OIST) Graduate University. We thank A. Chepelianskii and Y. Kubo for helpful discussions, and V.P. Dvornichenko for technical support.
\end{acknowledgments}

\clearpage
\widetext
\begin{center}
\textbf{\large Supplemental Material for ``Strong coupling of the cyclotron motion of surface electrons on liquid helium to a microwave cavity''}
\end{center}

\section{Fabry-Perot Resonator with 2DES}

\begin{figure}[b]
\begin{center}
\includegraphics[width=9cm,keepaspectratio]{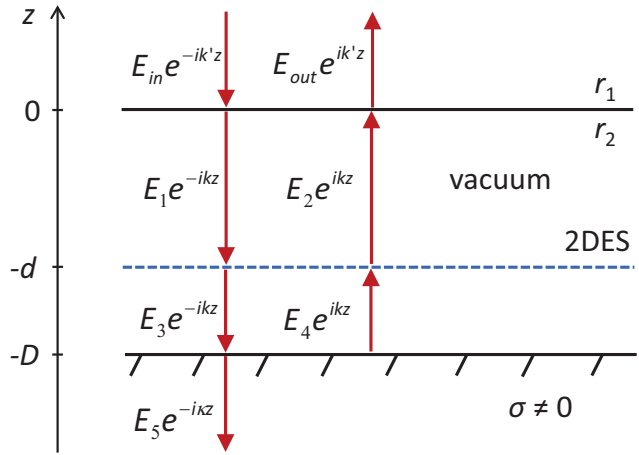}
\end{center}
\caption{\label{fig:1} (color online) Model for the Fabri-Perot resonator containing 2DES.}
\end{figure}

In our model, the Fabry-Perot Resonator (FPR) is formed by two infinitely large mirrors parallel to $xy$-plane located at distance $D$ apart from each other, see Fig.~\ref{fig:1}. The top mirror (at $z=0$) is a partially-reflecting mirror with the (amplitude) reflection coefficients $r_1$ and $r_2$ for the wave incident on the mirror from the top and bottom, respectively. The corresponding transmission coefficients are given by $t_i=1+r_i$, $i=1,2$. Here, we do not discuss physical realization of such a mirror. A trivial example is the interface between dielectric media with large dielectric constant (occupying the half-space $z>0$) and vacuum. The bottom mirror (occupying the half-space $z<-D$) is a good conductor with finite electrical conductivity $\sigma$, which accounts for internal losses of FPR. The infinitely large 2D Electron System (2DES) is located in $xy$ plane at distance $d<D$ below the top mirror . The problem of electro-magnetic (EM) field distribution inside and outside the resonator is solved by considering the superposition of propagating (in $z$-direction) EM waves and accounting for the boundary conditions at $z=-d$ and $-D$. We have
\begin{eqnarray}
&& E_1 = t_1 E_{in} + r_2 E_2, \nonumber \\
&& E_{out} =r_1 E_{in} + t_2 E_2, \nonumber \\
&& E_1 e^{ikd} + E_2 e^{-ikd} = E_3 e^{ikd} + E_4 e^{-ikd}, \nonumber \\
&& - \frac{E_3}{\eta_0} e^{ikd} + \frac{E_4}{\eta_0} e^{-ikd} - \left( - \frac{E_1}{\eta_0} e^{ikd} + \frac{E_2}{\eta_0} e^{-ikd} \right)  = j_x, \nonumber \\
&& E_3 e^{ikD} + E_4 e^{-ikD} = E_5 e^{i\kappa D}, \nonumber \\
&& -\frac{E_3}{\eta_0} e^{ikD} + \frac{E_4}{\eta_0} e^{-ikD} = -\frac{E_5}{\eta} e^{i\kappa D}.
\label{eq:sys}
\end{eqnarray} 
\noindent Here, the symbol $E$ stands for the phasor of linearly-polarized ($x$-direction) electric field, $j_x$ is the current density in 2DES, $\eta_0=\sqrt{\mu_0/\varepsilon_0}$ (377~Ohm) is the intrinsic impedance of vacuum, $k=\omega/c=\omega\sqrt{\varepsilon_0\mu_0}$ is the propagation constant in vacuum, $\kappa = \sqrt{\mu_0}\omega/\eta$ is the propagation constant in conductor, and $\eta$ is the intrinsic impedance of conductor:
\begin{equation}
\eta \approx \sqrt{\frac{\omega\mu_0}{2\sigma}} (1-i), \quad \frac{1}{\eta_0}\sqrt{\frac{\omega\mu_0}{2\sigma}}<<1.
\label{eq:aproxeta}
\end{equation}    
\noindent The third and forth equations in Eqs.(\ref{eq:sys}) express continuity of electric field and discontinuity of magnetic field, respectively, at $z=-d$. The latter is due to non-zero electric surface current in 2DES. The fifth and sixth equations express continuity of electric and magnetic fields, respectively, at $z=-D$. For the sake of simplicity, we assume that the dielectric constant of liquid helium is equal to 1.  

\indent From Eqs.~(\ref{eq:sys}) we obtain a relation between the electric field in the cavity at $z=-d$, that is $E_x=E_1 e^{ikd} + E_2 e^{-ikd}$, and the current density $j_x$. Arithmetics is significantly simplified  if we consider the frequency $\omega$ being close to $\omega_0=c\pi (m+1)/D$, where the mode number $m=0,1,..$ . In addition, we consider that 2DES is located at distance $\lambda_0/4=c\pi/(2\omega_0)$ from the second mirror, that is at the antinode of the electric field. Finally, we assume that $r_1\approx 1$ (that is $t_1\approx 2$) and $r_2\approx -1$ (that is $t_2<<1$). Expanding in the first order of $(\omega-\omega_0)/\omega_0$, $\sqrt{\omega\mu_0/(2\sigma)}/\eta_0$, and $t_2$, it is straightforward to obtain
\begin{equation}
\frac{D}{c} \Big( i(\omega - \omega_r) - (\gamma_{int}+\gamma_{ext})\Big) E_x - \eta_0 j_x = 2i(-1)^{(m+1)} E_{in}.
\label{eq:first}
\end{equation}
\noindent where $\omega_r=\omega_0 - \delta\omega_{int} - \delta\omega_{ext}$ is the resonant frequency of FPR, and 
\begin{equation}
\delta\omega_{int} = \frac{\omega_0}{\pi m}\sqrt{\frac{\omega\varepsilon_0}{2\sigma}}, \quad \delta\omega_{ext} = - \textrm{Im}\left( \frac{\omega_0}{2\pi (m+1)} t_2 \right), \quad \gamma_{int} = \frac{\omega_0}{\pi m}\sqrt{\frac{\omega\varepsilon_0}{2\sigma}}, \quad \gamma_{ext}=\textrm{Re}\left( \frac{\omega_0}{2\pi (m+1)} t_2 \right).
\label{eq:defenitions}
\end{equation}
\noindent Here, $\delta\omega_{int}$ and $\delta\omega_{ext}$ represent resonant frequency shifts, and $\gamma_{int}$ and $\gamma_{ext}$ represent internal and external loss rates of FPR, respectively,  

\section{AC conductivity of 2DES}

The second relation between the cavity field $E_x$ and current density $j_x$ can be written in terms of the longitudinal conductivity $\sigma_{xx}$, that is $j_x=\sigma_{xx} E_x$. The expression for AC conductivity tensor for 2D electrons in perpendicular magnetic field $B$ ($z$-direction) can be obtained, for example, in the memory function formalism~\cite{Monarkha2000}
\begin{equation}
\sigma_{xx} \pm i\sigma_{xy} = \frac{i n_e e^2}{m_e} \cdot \frac{1}{\omega \mp \omega_c + M(\omega)},
\end{equation}
\noindent where $\omega_c=eB/m_e$ is the cyclotron frequency, $m_e$ is the electron mass, and $n_e$ is the electron areal density. The real part of $M(\omega)$ determines the shift of the cyclotron resonance (CR) (usually negligibly small), while the imaginary part of $M(\omega)$ describes the CR linewidth. Defining $\nu=\textrm{Im}(M(\omega))$ and neglecting $\textrm{Re}(M(\omega))$, we obtain
\begin{equation}
\sigma_{xx} = \frac{n_e e^2}{m_e} \cdot \frac{i\omega - \nu}{(\omega + i\nu)^2 - \omega_c^2}.
\label{eq:sigma}
\end{equation}  

\section{Normal modes of coupled system of cavity field and 2DES motion}

\begin{figure}[t]
\begin{center}
\includegraphics[width=18cm,keepaspectratio]{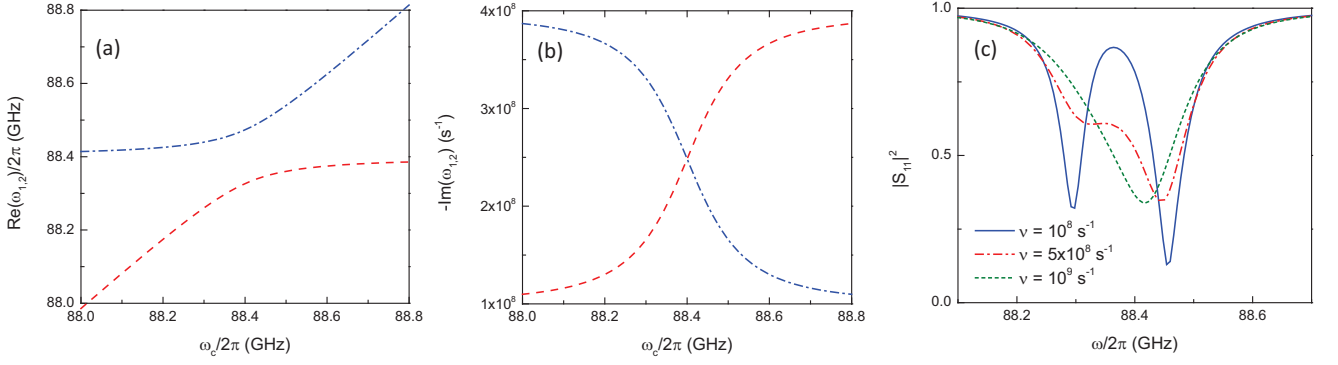}
\end{center}
\caption{\label{fig:2} (color online) (a) Real part of eigen frequencies $\omega_1$ (dash-dotted line, blue) and $\omega_2$ (dashed line, red)  of coupled system of cavity field and 2DES motion. (b) Imaginary part (with inverse sign) of eigen frequencies $\omega_1$ (dash-dotted line, blue) and $\omega_2$ (dashed line, red). Both graphs are calculated using Eq.~(\ref{eq:deter}) for $n_e=10^8$~cm$^{-2}$, $\nu=10^8$~s$^{-1}$, $\omega_r/2\pi=88.4$~GHz, $\gamma_{int}+\gamma_{ext}=4\times 10^8$~$s^{-1}$. (c) $|S_{11}|^2$ calculated using Eq.~(\ref{eq:S11}) for $n_e=10^8$~cm$^{-2}$, $\omega_c/2\pi=88.35$~GHz, $\omega_r/2\pi=88.4$~GHz, $\gamma_{int}+\gamma_{ext}=4\times 10^8$~$s^{-1}$, and several values of $\nu=10^8$ (solid line, blue), $5\times 10^8$ (dash-dotted line, red), and $10^9$~s$^{-1}$ (dashed line, green).}
\end{figure}

From Eqs. (\ref{eq:first}) and (\ref{eq:sigma}) we obtain a system of coupled equations (we consider the mode with $m=3$)
\begin{align}
\begin{pmatrix}
\frac{D}{2c} \Big[ (\omega - \omega_r) + i (\gamma_{int}+\gamma_{ext})\Big] \hfill &  \hfill i\eta_0/2 \\
- n_ee^2(i\omega - \nu)/m_e & \hfill (\omega + i\nu)^2 - \omega_c^2 \hfill
\end{pmatrix}
\begin{pmatrix}
E_x \\
j_x 
\end{pmatrix}
=
\begin{pmatrix}
 E_{in} \\
0 
\end{pmatrix}
\label{eq:coupledSM}
\end{align}

\noindent In the absence of external drive, $E_{in}=0$, the nontrivial solutions for $E_x$ and $j_x$ exist only for $\omega$ such that
\begin{align}
\textrm{det}
\begin{pmatrix}
(\omega - \omega_r) + i (\gamma_{int}+\gamma_{ext}) \hfill &  \hfill i/(\varepsilon_0 D) \\
- n_ee^2(i\omega - \nu)/m_e & \hfill (\omega + i\nu)^2 - \omega_c^2 \hfill
\end{pmatrix}
=0.
\label{eq:deter}
\end{align}

\noindent Eq.~(\ref{eq:deter}) provides frequencies $\omega_{1,2}$ for eigen modes of the coupled cavity-field-2DES motion. The real and imaginary parts of $\omega_{1,2}$ versus $\omega_c$ obtained from (\ref{eq:deter}) for $n_e=10^8$~cm$^{-2}$, $\nu=10^8$~s$^{-1}$, $\omega_r/2\pi=88.4$~GHz, and $\gamma_{int}+\gamma_{ext}=4\times 10^8$~$s^{-1}$ are shown in Fig.~\ref{fig:2}a,b. Two damped eigen modes exhibit splitting at $\omega_c = \omega_r$. It is instructive to find this splitting for the case of zero losses, that is $\nu=0$ and $\gamma_{int}+\gamma_{ext}=0$. In this case, Eq.~(\ref{eq:deter}) reduces to
\begin{equation}
-(\omega^2 - \omega_r^2)(\omega - \omega_r) + \frac{n_ee^2}{m_e\varepsilon_0 D} \omega = 0,
\end{equation}
\noindent which for $\omega$ close to $\omega_r$ has two solutions $\omega_{1,2}=\omega_r \pm g$, where $g=\sqrt{\frac{n_e e^2}{2m_e\varepsilon D}}$ is the coupling constant.

We note that this result is equivalent to the expression for $g$ in terms of the vacuum RMS electric field $E_0=\sqrt{\hbar\omega/2\varepsilon_0 V}$. Indeed, by multiplying and dividing the above classical result by $\hbar\omega_c$, and using $D=V/S$, where $S$ is the surface area of a finite-sized 2DES, we obtain
\begin{equation}
g=\sqrt{\frac{n_e e^2}{2m_e\varepsilon_0 D}}=\frac{e}{\hbar}\sqrt{\frac{\hbar}{m_e \omega_c}}\sqrt{\frac{\hbar\omega_c}{2\varepsilon_0 V}}\sqrt{n_e S}\approx \frac{el_BE_0}{\hbar}\sqrt{N_e},
\end{equation}
\noindent where $l_B=\sqrt{\hbar/m_e\omega_c}$ is the magnetic length, $N_e$ is the total number of electrons, and we used approximation $\omega \approx \omega_c$.  Also, note that since $D=c\pi (m+1)/\omega_0$, the coupling constant can be represented in the form $g\propto \sqrt{\alpha n_e}$, where $\alpha=e^2/(c\hbar)$ is the fine structure constant.

\indent For the sake of comparison with the experiment, it is also helpful to derive expression for the reflection parameter $S_{11}=E_{out}/E_{in}$. From Eqs.~(\ref{eq:sys}) it is straightforward to obtain
\begin{equation}
S_{11}= 1 + \frac{ 2(\gamma_{ext} - i\delta\omega_{ext}) }{ i(\omega -\omega_r) - (\gamma_{int}+\gamma_{ext}) - \frac{\sigma_{xx}}{\varepsilon_0 D}}.
\label{eq:S11}
\end{equation}  

\indent The numerical solutions for $|S_{11}|^2$ and $j_x/E_{in}$ obtained using Eqs.~(\ref{eq:coupledSM},\ref{eq:S11}) for the same conditions as in Fig.~\ref{fig:2}a,b are shown and compared with experimental results in Fig.~2 in the main text. 

\section{Normal-mode splitting versus damping rate}

The normal-mode splitting, which in the experiment appears in the cavity reflection spectrum, shows strong dependence on the damping rate $\nu$. Fig.~\ref{fig:2}c shows $|S_{11}|^2$ calculated using Eq.~(\ref{eq:S11}) for $n_e=10^8$~cm$^{-2}$, $\omega_c/2\pi=88.35$~GHz, $\omega_r/2\pi=88.4$~GHz, $\gamma_{int}+\gamma_{ext}=4\times 10^8$~$s^{-1}$, and several values of $\nu$. The splitting disappears for damping rates $\nu\gtrsim 10^9$~s$^{-1}$. This result should be compared with the experimental result shown in Fig.~3a in the main text. 


\begin{thebibliography}{40}%
\makeatletter
\providecommand \@ifxundefined [1]{%
 \@ifx{#1\undefined}
}%
\providecommand \@ifnum [1]{%
 \ifnum #1\expandafter \@firstoftwo
 \else \expandafter \@secondoftwo
 \fi
}%
\providecommand \@ifx [1]{%
 \ifx #1\expandafter \@firstoftwo
 \else \expandafter \@secondoftwo
 \fi
}%
\providecommand \natexlab [1]{#1}%
\providecommand \enquote  [1]{``#1''}%
\providecommand \bibnamefont  [1]{#1}%
\providecommand \bibfnamefont [1]{#1}%
\providecommand \citenamefont [1]{#1}%
\providecommand \href@noop [0]{\@secondoftwo}%
\providecommand \href [0]{\begingroup \@sanitize@url \@href}%
\providecommand \@href[1]{\@@startlink{#1}\@@href}%
\providecommand \@@href[1]{\endgroup#1\@@endlink}%
\providecommand \@sanitize@url [0]{\catcode `\\12\catcode `\$12\catcode
  `\&12\catcode `\#12\catcode `\^12\catcode `\_12\catcode `\%12\relax}%
\providecommand \@@startlink[1]{}%
\providecommand \@@endlink[0]{}%
\providecommand \url  [0]{\begingroup\@sanitize@url \@url }%
\providecommand \@url [1]{\endgroup\@href {#1}{\urlprefix }}%
\providecommand \urlprefix  [0]{URL }%
\providecommand \Eprint [0]{\href }%
\providecommand \doibase [0]{http://dx.doi.org/}%
\providecommand \selectlanguage [0]{\@gobble}%
\providecommand \bibinfo  [0]{\@secondoftwo}%
\providecommand \bibfield  [0]{\@secondoftwo}%
\providecommand \translation [1]{[#1]}%
\providecommand \BibitemOpen [0]{}%
\providecommand \bibitemStop [0]{}%
\providecommand \bibitemNoStop [0]{.\EOS\space}%
\providecommand \EOS [0]{\spacefactor3000\relax}%
\providecommand \BibitemShut  [1]{\csname bibitem#1\endcsname}%
\let\auto@bib@innerbib\@empty
%</preamble>
\bibitem [{\citenamefont {Kaluzny}\ \emph {et~al.}(1983)\citenamefont
  {Kaluzny}, \citenamefont {Goy}, \citenamefont {Gross}, \citenamefont
  {Raimond},\ and\ \citenamefont {Haroche}}]{Kaluzny1983}%
  \BibitemOpen
  \bibfield  {author} {\bibinfo {author} {\bibfnamefont {Y.}~\bibnamefont
  {Kaluzny}}, \bibinfo {author} {\bibfnamefont {P.}~\bibnamefont {Goy}},
  \bibinfo {author} {\bibfnamefont {M.}~\bibnamefont {Gross}}, \bibinfo
  {author} {\bibfnamefont {J.~M.}\ \bibnamefont {Raimond}}, \ and\ \bibinfo
  {author} {\bibfnamefont {S.}~\bibnamefont {Haroche}},\ }\href {\doibase
  10.1103/PhysRevLett.51.1175} {\bibfield  {journal} {\bibinfo  {journal}
  {Phys. Rev. Lett.}\ }\textbf {\bibinfo {volume} {51}},\ \bibinfo {pages}
  {1175} (\bibinfo {year} {1983})}\BibitemShut {NoStop}%
\bibitem [{\citenamefont {Zhu}\ \emph {et~al.}(1990)\citenamefont {Zhu},
  \citenamefont {Gauthier}, \citenamefont {Morin}, \citenamefont {Wu},
  \citenamefont {Carmichael},\ and\ \citenamefont {Mossberg}}]{Zhu1990}%
  \BibitemOpen
  \bibfield  {author} {\bibinfo {author} {\bibfnamefont {Y.}~\bibnamefont
  {Zhu}}, \bibinfo {author} {\bibfnamefont {D.~J.}\ \bibnamefont {Gauthier}},
  \bibinfo {author} {\bibfnamefont {S.~E.}\ \bibnamefont {Morin}}, \bibinfo
  {author} {\bibfnamefont {Q.}~\bibnamefont {Wu}}, \bibinfo {author}
  {\bibfnamefont {H.~J.}\ \bibnamefont {Carmichael}}, \ and\ \bibinfo {author}
  {\bibfnamefont {T.~W.}\ \bibnamefont {Mossberg}},\ }\href {\doibase
  10.1103/PhysRevLett.64.2499} {\bibfield  {journal} {\bibinfo  {journal}
  {Phys. Rev. Lett.}\ }\textbf {\bibinfo {volume} {64}},\ \bibinfo {pages}
  {2499} (\bibinfo {year} {1990})}\BibitemShut {NoStop}%
\bibitem [{\citenamefont {Thompson}\ \emph {et~al.}(1992)\citenamefont
  {Thompson}, \citenamefont {Rempe},\ and\ \citenamefont
  {Kimble}}]{Thompson1992}%
  \BibitemOpen
  \bibfield  {author} {\bibinfo {author} {\bibfnamefont {R.~J.}\ \bibnamefont
  {Thompson}}, \bibinfo {author} {\bibfnamefont {G.}~\bibnamefont {Rempe}}, \
  and\ \bibinfo {author} {\bibfnamefont {H.~J.}\ \bibnamefont {Kimble}},\
  }\href {\doibase 10.1103/PhysRevLett.68.1132} {\bibfield  {journal} {\bibinfo
   {journal} {Phys. Rev. Lett.}\ }\textbf {\bibinfo {volume} {68}},\ \bibinfo
  {pages} {1132} (\bibinfo {year} {1992})}\BibitemShut {NoStop}%
\bibitem [{\citenamefont {Chiorescu}\ \emph {et~al.}(2010)\citenamefont
  {Chiorescu}, \citenamefont {Groll}, \citenamefont {Bertaina}, \citenamefont
  {Mori},\ and\ \citenamefont {Miyashita}}]{Chiorescu2010}%
  \BibitemOpen
  \bibfield  {author} {\bibinfo {author} {\bibfnamefont {I.}~\bibnamefont
  {Chiorescu}}, \bibinfo {author} {\bibfnamefont {N.}~\bibnamefont {Groll}},
  \bibinfo {author} {\bibfnamefont {S.}~\bibnamefont {Bertaina}}, \bibinfo
  {author} {\bibfnamefont {T.}~\bibnamefont {Mori}}, \ and\ \bibinfo {author}
  {\bibfnamefont {S.}~\bibnamefont {Miyashita}},\ }\href {\doibase
  10.1103/PhysRevB.82.024413} {\bibfield  {journal} {\bibinfo  {journal} {Phys.
  Rev. B}\ }\textbf {\bibinfo {volume} {82}},\ \bibinfo {pages} {024413}
  (\bibinfo {year} {2010})}\BibitemShut {NoStop}%
\bibitem [{\citenamefont {Schuster}\ \emph {et~al.}(2010)\citenamefont
  {Schuster}, \citenamefont {Sears}, \citenamefont {Ginossar}, \citenamefont
  {DiCarlo}, \citenamefont {Frunzio}, \citenamefont {Morton}, \citenamefont
  {Wu}, \citenamefont {Briggs}, \citenamefont {Buckley}, \citenamefont
  {Awschalom},\ and\ \citenamefont {Schoelkopf}}]{Schuster2010}%
  \BibitemOpen
  \bibfield  {author} {\bibinfo {author} {\bibfnamefont {D.~I.}\ \bibnamefont
  {Schuster}}, \bibinfo {author} {\bibfnamefont {A.~P.}\ \bibnamefont {Sears}},
  \bibinfo {author} {\bibfnamefont {E.}~\bibnamefont {Ginossar}}, \bibinfo
  {author} {\bibfnamefont {L.}~\bibnamefont {DiCarlo}}, \bibinfo {author}
  {\bibfnamefont {L.}~\bibnamefont {Frunzio}}, \bibinfo {author} {\bibfnamefont
  {J.~J.~L.}\ \bibnamefont {Morton}}, \bibinfo {author} {\bibfnamefont
  {H.}~\bibnamefont {Wu}}, \bibinfo {author} {\bibfnamefont {G.~A.~D.}\
  \bibnamefont {Briggs}}, \bibinfo {author} {\bibfnamefont {B.~B.}\
  \bibnamefont {Buckley}}, \bibinfo {author} {\bibfnamefont {D.~D.}\
  \bibnamefont {Awschalom}}, \ and\ \bibinfo {author} {\bibfnamefont {R.~J.}\
  \bibnamefont {Schoelkopf}},\ }\href {\doibase 10.1103/PhysRevLett.105.140501}
  {\bibfield  {journal} {\bibinfo  {journal} {Phys. Rev. Lett.}\ }\textbf
  {\bibinfo {volume} {105}},\ \bibinfo {pages} {140501} (\bibinfo {year}
  {2010})}\BibitemShut {NoStop}%
\bibitem [{\citenamefont {Kubo}\ \emph {et~al.}(2010)\citenamefont {Kubo},
  \citenamefont {Ong}, \citenamefont {Bertet}, \citenamefont {Vion},
  \citenamefont {Jacques}, \citenamefont {Zheng}, \citenamefont {Dreau},
  \citenamefont {Roch}, \citenamefont {Auffeves}, \citenamefont {Jelezko},
  \citenamefont {Wrachtrup}, \citenamefont {Barthe}, \citenamefont {Bergonzo},\
  and\ \citenamefont {Esteve}}]{Kubo2010}%
  \BibitemOpen
  \bibfield  {author} {\bibinfo {author} {\bibfnamefont {Y.}~\bibnamefont
  {Kubo}}, \bibinfo {author} {\bibfnamefont {F.~R.}\ \bibnamefont {Ong}},
  \bibinfo {author} {\bibfnamefont {P.}~\bibnamefont {Bertet}}, \bibinfo
  {author} {\bibfnamefont {D.}~\bibnamefont {Vion}}, \bibinfo {author}
  {\bibfnamefont {V.}~\bibnamefont {Jacques}}, \bibinfo {author} {\bibfnamefont
  {D.}~\bibnamefont {Zheng}}, \bibinfo {author} {\bibfnamefont
  {A.}~\bibnamefont {Dreau}}, \bibinfo {author} {\bibfnamefont {J.-F.}\
  \bibnamefont {Roch}}, \bibinfo {author} {\bibfnamefont {A.}~\bibnamefont
  {Auffeves}}, \bibinfo {author} {\bibfnamefont {F.}~\bibnamefont {Jelezko}},
  \bibinfo {author} {\bibfnamefont {J.}~\bibnamefont {Wrachtrup}}, \bibinfo
  {author} {\bibfnamefont {M.~F.}\ \bibnamefont {Barthe}}, \bibinfo {author}
  {\bibfnamefont {P.}~\bibnamefont {Bergonzo}}, \ and\ \bibinfo {author}
  {\bibfnamefont {D.}~\bibnamefont {Esteve}},\ }\href {\doibase
  10.1103/PhysRevLett.105.140502} {\bibfield  {journal} {\bibinfo  {journal}
  {Phys. Rev. Lett.}\ }\textbf {\bibinfo {volume} {105}},\ \bibinfo {pages}
  {140502} (\bibinfo {year} {2010})}\BibitemShut {NoStop}%
\bibitem [{\citenamefont {Amsuss}\ \emph {et~al.}(2011)\citenamefont {Amsuss},
  \citenamefont {Koller}, \citenamefont {Nobauer}, \citenamefont {Putz},
  \citenamefont {Rotter}, \citenamefont {Sandner}, \citenamefont {Schneider},
  \citenamefont {Schrambock}, \citenamefont {Steinhauser}, \citenamefont
  {Ritsch}, \citenamefont {Schmiedmayer},\ and\ \citenamefont
  {Majer}}]{Amsuss2011}%
  \BibitemOpen
  \bibfield  {author} {\bibinfo {author} {\bibfnamefont {R.}~\bibnamefont
  {Amsuss}}, \bibinfo {author} {\bibfnamefont {C.}~\bibnamefont {Koller}},
  \bibinfo {author} {\bibfnamefont {T.}~\bibnamefont {Nobauer}}, \bibinfo
  {author} {\bibfnamefont {S.}~\bibnamefont {Putz}}, \bibinfo {author}
  {\bibfnamefont {S.}~\bibnamefont {Rotter}}, \bibinfo {author} {\bibfnamefont
  {K.}~\bibnamefont {Sandner}}, \bibinfo {author} {\bibfnamefont
  {S.}~\bibnamefont {Schneider}}, \bibinfo {author} {\bibfnamefont
  {M.}~\bibnamefont {Schrambock}}, \bibinfo {author} {\bibfnamefont
  {G.}~\bibnamefont {Steinhauser}}, \bibinfo {author} {\bibfnamefont
  {H.}~\bibnamefont {Ritsch}}, \bibinfo {author} {\bibfnamefont
  {J.}~\bibnamefont {Schmiedmayer}}, \ and\ \bibinfo {author} {\bibfnamefont
  {J.}~\bibnamefont {Majer}},\ }\href {\doibase 10.1103/PhysRevLett.107.060502}
  {\bibfield  {journal} {\bibinfo  {journal} {Phys. Rev. Lett.}\ }\textbf
  {\bibinfo {volume} {107}},\ \bibinfo {pages} {060502} (\bibinfo {year}
  {2011})}\BibitemShut {NoStop}%
\bibitem [{\citenamefont {Abe}\ \emph {et~al.}(2011)\citenamefont {Abe},
  \citenamefont {Wu}, \citenamefont {Ardavan},\ and\ \citenamefont
  {Morton}}]{Abe2011}%
  \BibitemOpen
  \bibfield  {author} {\bibinfo {author} {\bibfnamefont {E.}~\bibnamefont
  {Abe}}, \bibinfo {author} {\bibfnamefont {H.}~\bibnamefont {Wu}}, \bibinfo
  {author} {\bibfnamefont {A.}~\bibnamefont {Ardavan}}, \ and\ \bibinfo
  {author} {\bibfnamefont {J.~J.~L.}\ \bibnamefont {Morton}},\ }\href {\doibase
  10.1063/1.3601930} {\bibfield  {journal} {\bibinfo  {journal} {Appl. Phys.
  Lett.}\ }\textbf {\bibinfo {volume} {98}},\ \bibinfo {pages} {251108}
  (\bibinfo {year} {2011})}\BibitemShut {NoStop}%
\bibitem [{\citenamefont {Huebl}\ \emph {et~al.}(2013)\citenamefont {Huebl},
  \citenamefont {Zollitsch}, \citenamefont {Lotze}, \citenamefont {Hocke},
  \citenamefont {Greifenstein}, \citenamefont {Marx}, \citenamefont {Gross},\
  and\ \citenamefont {Goennenwein}}]{Huebl2013}%
  \BibitemOpen
  \bibfield  {author} {\bibinfo {author} {\bibfnamefont {H.}~\bibnamefont
  {Huebl}}, \bibinfo {author} {\bibfnamefont {C.~W.}\ \bibnamefont
  {Zollitsch}}, \bibinfo {author} {\bibfnamefont {J.}~\bibnamefont {Lotze}},
  \bibinfo {author} {\bibfnamefont {F.}~\bibnamefont {Hocke}}, \bibinfo
  {author} {\bibfnamefont {M.}~\bibnamefont {Greifenstein}}, \bibinfo {author}
  {\bibfnamefont {A.}~\bibnamefont {Marx}}, \bibinfo {author} {\bibfnamefont
  {R.}~\bibnamefont {Gross}}, \ and\ \bibinfo {author} {\bibfnamefont
  {S.~T.~B.}\ \bibnamefont {Goennenwein}},\ }\href {\doibase
  10.1103/PhysRevLett.111.127003} {\bibfield  {journal} {\bibinfo  {journal}
  {Phys. Rev. Lett.}\ }\textbf {\bibinfo {volume} {111}},\ \bibinfo {pages}
  {127003} (\bibinfo {year} {2013})}\BibitemShut {NoStop}%
\bibitem [{\citenamefont {Tabuchi}\ \emph {et~al.}(2014)\citenamefont
  {Tabuchi}, \citenamefont {Ishino}, \citenamefont {Ishikawa}, \citenamefont
  {Yamazaki}, \citenamefont {Usami},\ and\ \citenamefont
  {Nakamura}}]{Tabuchi2014}%
  \BibitemOpen
  \bibfield  {author} {\bibinfo {author} {\bibfnamefont {Y.}~\bibnamefont
  {Tabuchi}}, \bibinfo {author} {\bibfnamefont {S.}~\bibnamefont {Ishino}},
  \bibinfo {author} {\bibfnamefont {T.}~\bibnamefont {Ishikawa}}, \bibinfo
  {author} {\bibfnamefont {R.}~\bibnamefont {Yamazaki}}, \bibinfo {author}
  {\bibfnamefont {K.}~\bibnamefont {Usami}}, \ and\ \bibinfo {author}
  {\bibfnamefont {Y.}~\bibnamefont {Nakamura}},\ }\href {\doibase
  10.1103/PhysRevLett.113.083603} {\bibfield  {journal} {\bibinfo  {journal}
  {Phys. Rev. Lett.}\ }\textbf {\bibinfo {volume} {113}},\ \bibinfo {pages}
  {083603} (\bibinfo {year} {2014})}\BibitemShut {NoStop}%
\bibitem [{\citenamefont {Zhang}\ \emph {et~al.}(2014)\citenamefont {Zhang},
  \citenamefont {Zou}, \citenamefont {Jiang},\ and\ \citenamefont
  {Tang}}]{Zhang2014}%
  \BibitemOpen
  \bibfield  {author} {\bibinfo {author} {\bibfnamefont {X.}~\bibnamefont
  {Zhang}}, \bibinfo {author} {\bibfnamefont {C.-L.}\ \bibnamefont {Zou}},
  \bibinfo {author} {\bibfnamefont {L.}~\bibnamefont {Jiang}}, \ and\ \bibinfo
  {author} {\bibfnamefont {H.~X.}\ \bibnamefont {Tang}},\ }\href {\doibase
  10.1103/PhysRevLett.113.156401} {\bibfield  {journal} {\bibinfo  {journal}
  {Phys. Rev. Lett.}\ }\textbf {\bibinfo {volume} {113}},\ \bibinfo {pages}
  {156401} (\bibinfo {year} {2014})}\BibitemShut {NoStop}%
\bibitem [{\citenamefont {Abdurakhimov}\ \emph {et~al.}(2015)\citenamefont
  {Abdurakhimov}, \citenamefont {Bunkov},\ and\ \citenamefont
  {Konstantinov}}]{Abdurakhimov2015}%
  \BibitemOpen
  \bibfield  {author} {\bibinfo {author} {\bibfnamefont {L.~V.}\ \bibnamefont
  {Abdurakhimov}}, \bibinfo {author} {\bibfnamefont {Y.~M.}\ \bibnamefont
  {Bunkov}}, \ and\ \bibinfo {author} {\bibfnamefont {D.}~\bibnamefont
  {Konstantinov}},\ }\href {\doibase 10.1103/PhysRevLett.114.226402} {\bibfield
   {journal} {\bibinfo  {journal} {Phys. Rev. Lett.}\ }\textbf {\bibinfo
  {volume} {114}},\ \bibinfo {pages} {226402} (\bibinfo {year}
  {2015})}\BibitemShut {NoStop}%
\bibitem [{\citenamefont {Hagenmuller}\ \emph {et~al.}(2010)\citenamefont
  {Hagenmuller}, \citenamefont {De~Liberato},\ and\ \citenamefont
  {Ciuti}}]{Hagenmuller2010}%
  \BibitemOpen
  \bibfield  {author} {\bibinfo {author} {\bibfnamefont {D.}~\bibnamefont
  {Hagenmuller}}, \bibinfo {author} {\bibfnamefont {S.}~\bibnamefont
  {De~Liberato}}, \ and\ \bibinfo {author} {\bibfnamefont {C.}~\bibnamefont
  {Ciuti}},\ }\href {\doibase 10.1103/PhysRevB.81.235303} {\bibfield  {journal}
  {\bibinfo  {journal} {Phys. Rev. B}\ }\textbf {\bibinfo {volume} {81}},\
  \bibinfo {pages} {235303} (\bibinfo {year} {2010})}\BibitemShut {NoStop}%
\bibitem [{\citenamefont {Scalari}\ \emph {et~al.}(2012)\citenamefont
  {Scalari}, \citenamefont {Maissen}, \citenamefont {Tur\u{c}inkov\`a},
  \citenamefont {Hagenm\"uller}, \citenamefont {De~Liberato}, \citenamefont
  {Ciuti}, \citenamefont {Reichl}, \citenamefont {Schuh}, \citenamefont
  {Wegscheider}, \citenamefont {Beck},\ and\ \citenamefont
  {Faist}}]{Scalari2012}%
  \BibitemOpen
  \bibfield  {author} {\bibinfo {author} {\bibfnamefont {G.}~\bibnamefont
  {Scalari}}, \bibinfo {author} {\bibfnamefont {C.}~\bibnamefont {Maissen}},
  \bibinfo {author} {\bibfnamefont {D.}~\bibnamefont {Tur\u{c}inkov\`a}},
  \bibinfo {author} {\bibfnamefont {D.}~\bibnamefont {Hagenm\"uller}}, \bibinfo
  {author} {\bibfnamefont {S.}~\bibnamefont {De~Liberato}}, \bibinfo {author}
  {\bibfnamefont {C.}~\bibnamefont {Ciuti}}, \bibinfo {author} {\bibfnamefont
  {C.}~\bibnamefont {Reichl}}, \bibinfo {author} {\bibfnamefont
  {D.}~\bibnamefont {Schuh}}, \bibinfo {author} {\bibfnamefont
  {W.}~\bibnamefont {Wegscheider}}, \bibinfo {author} {\bibfnamefont
  {M.}~\bibnamefont {Beck}}, \ and\ \bibinfo {author} {\bibfnamefont
  {J.}~\bibnamefont {Faist}},\ }\href {\doibase 10.1126/science.1216022}
  {\bibfield  {journal} {\bibinfo  {journal} {Science}\ }\textbf {\bibinfo
  {volume} {335}},\ \bibinfo {pages} {1323} (\bibinfo {year}
  {2012})}\BibitemShut {NoStop}%
\bibitem [{\citenamefont {Muravev}\ \emph {et~al.}(2011)\citenamefont
  {Muravev}, \citenamefont {Andreev}, \citenamefont {Kukushkin}, \citenamefont
  {Schmult},\ and\ \citenamefont {Dietsche}}]{Muravev2011}%
  \BibitemOpen
  \bibfield  {author} {\bibinfo {author} {\bibfnamefont {V.~M.}\ \bibnamefont
  {Muravev}}, \bibinfo {author} {\bibfnamefont {I.~V.}\ \bibnamefont
  {Andreev}}, \bibinfo {author} {\bibfnamefont {I.~V.}\ \bibnamefont
  {Kukushkin}}, \bibinfo {author} {\bibfnamefont {S.}~\bibnamefont {Schmult}},
  \ and\ \bibinfo {author} {\bibfnamefont {W.}~\bibnamefont {Dietsche}},\
  }\href {\doibase 10.1103/PhysRevB.83.075309} {\bibfield  {journal} {\bibinfo
  {journal} {Phys. Rev. B}\ }\textbf {\bibinfo {volume} {83}},\ \bibinfo
  {pages} {075309} (\bibinfo {year} {2011})}\BibitemShut {NoStop}%
\bibitem [{\citenamefont {Muravev}\ \emph {et~al.}(2013)\citenamefont
  {Muravev}, \citenamefont {Gusikhin}, \citenamefont {Andreev},\ and\
  \citenamefont {Kukushkin}}]{Muravev2013}%
  \BibitemOpen
  \bibfield  {author} {\bibinfo {author} {\bibfnamefont {V.~M.}\ \bibnamefont
  {Muravev}}, \bibinfo {author} {\bibfnamefont {P.~A.}\ \bibnamefont
  {Gusikhin}}, \bibinfo {author} {\bibfnamefont {I.~V.}\ \bibnamefont
  {Andreev}}, \ and\ \bibinfo {author} {\bibfnamefont {I.~V.}\ \bibnamefont
  {Kukushkin}},\ }\href {\doibase 10.1103/PhysRevB.87.045307} {\bibfield
  {journal} {\bibinfo  {journal} {Phys. Rev. B}\ }\textbf {\bibinfo {volume}
  {87}},\ \bibinfo {pages} {045307} (\bibinfo {year} {2013})}\BibitemShut
  {NoStop}%
\bibitem [{\citenamefont {Zhang}\ \emph {et~al.}()\citenamefont {Zhang},
  \citenamefont {Lou}, \citenamefont {Li}, \citenamefont {Reno}, \citenamefont
  {Pan}, \citenamefont {Watson}, \citenamefont {Manfra},\ and\ \citenamefont
  {Kono}}]{QiZhang2016}%
  \BibitemOpen
  \bibfield  {author} {\bibinfo {author} {\bibfnamefont {Q.}~\bibnamefont
  {Zhang}}, \bibinfo {author} {\bibfnamefont {M.}~\bibnamefont {Lou}}, \bibinfo
  {author} {\bibfnamefont {X.}~\bibnamefont {Li}}, \bibinfo {author}
  {\bibfnamefont {J.}~\bibnamefont {Reno}}, \bibinfo {author} {\bibfnamefont
  {W.}~\bibnamefont {Pan}}, \bibinfo {author} {\bibfnamefont {J.}~\bibnamefont
  {Watson}}, \bibinfo {author} {\bibfnamefont {M.}~\bibnamefont {Manfra}}, \
  and\ \bibinfo {author} {\bibfnamefont {J.}~\bibnamefont {Kono}},\ }\href@noop
  {} {}\Eprint {http://arxiv.org/abs/1604.08297} {arXiv:1604.08297}
  \BibitemShut {NoStop}%
\bibitem [{\citenamefont {Hagenmuller}\ and\ \citenamefont
  {Ciuti}(2012)}]{Hagenmuller2012}%
  \BibitemOpen
  \bibfield  {author} {\bibinfo {author} {\bibfnamefont {D.}~\bibnamefont
  {Hagenmuller}}\ and\ \bibinfo {author} {\bibfnamefont {C.}~\bibnamefont
  {Ciuti}},\ }\href {\doibase 10.1103/PhysRevLett.109.267403} {\bibfield
  {journal} {\bibinfo  {journal} {Phys. Rev. Lett.}\ }\textbf {\bibinfo
  {volume} {109}},\ \bibinfo {pages} {267403} (\bibinfo {year}
  {2012})}\BibitemShut {NoStop}%
\bibitem [{\citenamefont {Chirolli}\ \emph {et~al.}(2012)\citenamefont
  {Chirolli}, \citenamefont {Polini}, \citenamefont {Giovannetti},\ and\
  \citenamefont {MacDonald}}]{Chirolli2012}%
  \BibitemOpen
  \bibfield  {author} {\bibinfo {author} {\bibfnamefont {L.}~\bibnamefont
  {Chirolli}}, \bibinfo {author} {\bibfnamefont {M.}~\bibnamefont {Polini}},
  \bibinfo {author} {\bibfnamefont {V.}~\bibnamefont {Giovannetti}}, \ and\
  \bibinfo {author} {\bibfnamefont {A.~H.}\ \bibnamefont {MacDonald}},\ }\href
  {\doibase 10.1103/PhysRevLett.109.267404} {\bibfield  {journal} {\bibinfo
  {journal} {Phys. Rev. Lett.}\ }\textbf {\bibinfo {volume} {109}},\ \bibinfo
  {pages} {267404} (\bibinfo {year} {2012})}\BibitemShut {NoStop}%
\bibitem [{\citenamefont {Agarwal}(1984)}]{Agarwal1984}%
  \BibitemOpen
  \bibfield  {author} {\bibinfo {author} {\bibfnamefont {G.~S.}\ \bibnamefont
  {Agarwal}},\ }\href {\doibase 10.1103/PhysRevLett.53.1732} {\bibfield
  {journal} {\bibinfo  {journal} {Phys. Rev. Lett.}\ }\textbf {\bibinfo
  {volume} {53}},\ \bibinfo {pages} {1732} (\bibinfo {year}
  {1984})}\BibitemShut {NoStop}%
\bibitem [{\citenamefont {Imamoglu}(2009)}]{Imamoglu2009}%
  \BibitemOpen
  \bibfield  {author} {\bibinfo {author} {\bibfnamefont {A.}~\bibnamefont
  {Imamoglu}},\ }\href {\doibase 10.1103/PhysRevLett.102.083602} {\bibfield
  {journal} {\bibinfo  {journal} {Phys. Rev. Lett.}\ }\textbf {\bibinfo
  {volume} {102}},\ \bibinfo {pages} {083602} (\bibinfo {year}
  {2009})}\BibitemShut {NoStop}%
\bibitem [{\citenamefont {Bai}\ \emph {et~al.}(2015)\citenamefont {Bai},
  \citenamefont {Harder}, \citenamefont {Chen}, \citenamefont {Fan},
  \citenamefont {Xiao},\ and\ \citenamefont {Hu}}]{Bai2015}%
  \BibitemOpen
  \bibfield  {author} {\bibinfo {author} {\bibfnamefont {L.}~\bibnamefont
  {Bai}}, \bibinfo {author} {\bibfnamefont {M.}~\bibnamefont {Harder}},
  \bibinfo {author} {\bibfnamefont {Y.~P.}\ \bibnamefont {Chen}}, \bibinfo
  {author} {\bibfnamefont {X.}~\bibnamefont {Fan}}, \bibinfo {author}
  {\bibfnamefont {J.~Q.}\ \bibnamefont {Xiao}}, \ and\ \bibinfo {author}
  {\bibfnamefont {C.-M.}\ \bibnamefont {Hu}},\ }\href {\doibase
  10.1103/PhysRevLett.114.227201} {\bibfield  {journal} {\bibinfo  {journal}
  {Phys. Rev. Lett.}\ }\textbf {\bibinfo {volume} {114}},\ \bibinfo {pages}
  {227201} (\bibinfo {year} {2015})}\BibitemShut {NoStop}%
\bibitem [{\citenamefont {Hofheinz}\ \emph {et~al.}(2008)\citenamefont
  {Hofheinz}, \citenamefont {Weig}, \citenamefont {Ansmann}, \citenamefont
  {Bialczak}, \citenamefont {Lucero}, \citenamefont {Neeley}, \citenamefont
  {O'Connell}, \citenamefont {Wang}, \citenamefont {Martinis},\ and\
  \citenamefont {Cleland}}]{Hofheinz2008}%
  \BibitemOpen
  \bibfield  {author} {\bibinfo {author} {\bibfnamefont {M.}~\bibnamefont
  {Hofheinz}}, \bibinfo {author} {\bibfnamefont {E.~M.}\ \bibnamefont {Weig}},
  \bibinfo {author} {\bibfnamefont {M.}~\bibnamefont {Ansmann}}, \bibinfo
  {author} {\bibfnamefont {R.~C.}\ \bibnamefont {Bialczak}}, \bibinfo {author}
  {\bibfnamefont {E.}~\bibnamefont {Lucero}}, \bibinfo {author} {\bibfnamefont
  {M.}~\bibnamefont {Neeley}}, \bibinfo {author} {\bibfnamefont {A.~D.}\
  \bibnamefont {O'Connell}}, \bibinfo {author} {\bibfnamefont {H.}~\bibnamefont
  {Wang}}, \bibinfo {author} {\bibfnamefont {J.~M.}\ \bibnamefont {Martinis}},
  \ and\ \bibinfo {author} {\bibfnamefont {A.~N.}\ \bibnamefont {Cleland}},\
  }\href {\doibase 10.1038/nature07136} {\bibfield  {journal} {\bibinfo
  {journal} {Nature}\ }\textbf {\bibinfo {volume} {454}},\ \bibinfo {pages}
  {310} (\bibinfo {year} {2008})}\BibitemShut {NoStop}%
\bibitem [{\citenamefont {Haroche}\ and\ \citenamefont
  {Raimond}(2006)}]{HarocheRaimond}%
  \BibitemOpen
  \bibfield  {author} {\bibinfo {author} {\bibfnamefont {S.}~\bibnamefont
  {Haroche}}\ and\ \bibinfo {author} {\bibfnamefont {J.-M.}\ \bibnamefont
  {Raimond}},\ }\href@noop {} {\emph {\bibinfo {title} {Exploring the Quantum:
  Atoms, Cavities and Photons}}},\ \bibinfo {number} {Chapter 3}\ (\bibinfo
  {publisher} {Oxford University Press},\ \bibinfo {year} {2006})\BibitemShut
  {NoStop}%
\bibitem [{\citenamefont {Yamashiro}\ \emph {et~al.}(2015)\citenamefont
  {Yamashiro}, \citenamefont {Abdurakhimov}, \citenamefont {Badrutdinov},
  \citenamefont {Monarkha},\ and\ \citenamefont
  {Konstantinov}}]{Yamashiro2015}%
  \BibitemOpen
  \bibfield  {author} {\bibinfo {author} {\bibfnamefont {R.}~\bibnamefont
  {Yamashiro}}, \bibinfo {author} {\bibfnamefont {L.~V.}\ \bibnamefont
  {Abdurakhimov}}, \bibinfo {author} {\bibfnamefont {A.~O.}\ \bibnamefont
  {Badrutdinov}}, \bibinfo {author} {\bibfnamefont {Y.~P.}\ \bibnamefont
  {Monarkha}}, \ and\ \bibinfo {author} {\bibfnamefont {D.}~\bibnamefont
  {Konstantinov}},\ }\href {\doibase 10.1103/PhysRevLett.115.256802} {\bibfield
   {journal} {\bibinfo  {journal} {Phys. Rev. Lett.}\ }\textbf {\bibinfo
  {volume} {115}},\ \bibinfo {pages} {256802} (\bibinfo {year}
  {2015})}\BibitemShut {NoStop}%
\bibitem [{\citenamefont {Kogelnik}\ and\ \citenamefont
  {Li}(1966)}]{Kogelnik1966}%
  \BibitemOpen
  \bibfield  {author} {\bibinfo {author} {\bibfnamefont {H.}~\bibnamefont
  {Kogelnik}}\ and\ \bibinfo {author} {\bibfnamefont {T.}~\bibnamefont {Li}},\
  }\href {\doibase 10.1364/AO.5.001550} {\bibfield  {journal} {\bibinfo
  {journal} {Appl. Opt.}\ }\textbf {\bibinfo {volume} {5}},\ \bibinfo {pages}
  {1550} (\bibinfo {year} {1966})}\BibitemShut {NoStop}%
\bibitem [{TEM()}]{TEM_notation}%
  \BibitemOpen
  \href@noop {} {\emph {\bibinfo {title} {We use notation $TEM_{00q}$ with
  $q=0,1,2,\ldots$, where the number of half wavelengths is $q+1$. See details
  in~\cite{Kogelnik1966}}}}\BibitemShut {NoStop}%
\bibitem [{\citenamefont {Andrei}(2010)}]{Andrei}%
  \BibitemOpen
  \bibfield  {author} {\bibinfo {author} {\bibfnamefont {E.~Y.}\ \bibnamefont
  {Andrei}},\ }\href@noop {} {\emph {\bibinfo {title} {Two-Dimensional Electron
  Systems on Helium and other Cyrogenic Substrates}}},\ edited by\ \bibinfo
  {editor} {\bibfnamefont {E.~Y.}\ \bibnamefont {Andrei}}\ (\bibinfo
  {publisher} {Kluwer Academic Publishers},\ \bibinfo {year}
  {2010})\BibitemShut {NoStop}%
\bibitem [{\citenamefont {Monarkha}\ and\ \citenamefont
  {Kono}(2010)}]{MonarkhaKono}%
  \BibitemOpen
  \bibfield  {author} {\bibinfo {author} {\bibfnamefont {Y.}~\bibnamefont
  {Monarkha}}\ and\ \bibinfo {author} {\bibfnamefont {K.}~\bibnamefont
  {Kono}},\ }\href@noop {} {\emph {\bibinfo {title} {Two-Dimensional Coulomb
  Liquids and Solids}}}\ (\bibinfo  {publisher} {Springer},\ \bibinfo {year}
  {2010})\BibitemShut {NoStop}%
\bibitem [{\citenamefont {Edel'man}(1977)}]{Edelman1977}%
  \BibitemOpen
  \bibfield  {author} {\bibinfo {author} {\bibfnamefont {V.~S.}\ \bibnamefont
  {Edel'man}},\ }\href@noop {} {\bibfield  {journal} {\bibinfo  {journal} {JETP
  Letters}\ }\textbf {\bibinfo {volume} {25}},\ \bibinfo {pages} {394}
  (\bibinfo {year} {1977})}\BibitemShut {NoStop}%
\bibitem [{\citenamefont {Aoki}\ and\ \citenamefont {Saitoh}(1980)}]{Aoki1980}%
  \BibitemOpen
  \bibfield  {author} {\bibinfo {author} {\bibfnamefont {T.}~\bibnamefont
  {Aoki}}\ and\ \bibinfo {author} {\bibfnamefont {M.}~\bibnamefont {Saitoh}},\
  }\href {\doibase 10.1143/JPSJ.48.1929} {\bibfield  {journal} {\bibinfo
  {journal} {J. Phys. Soc. Jpn.}\ }\textbf {\bibinfo {volume} {48}},\ \bibinfo
  {pages} {1929} (\bibinfo {year} {1980})}\BibitemShut {NoStop}%
\bibitem [{\citenamefont {Mehrotra}\ and\ \citenamefont
  {Dahm}(1987)}]{Mehrotra1987}%
  \BibitemOpen
  \bibfield  {author} {\bibinfo {author} {\bibfnamefont {R.}~\bibnamefont
  {Mehrotra}}\ and\ \bibinfo {author} {\bibfnamefont {A.}~\bibnamefont
  {Dahm}},\ }\href {\doibase 10.1007/BF01070654} {\bibfield  {journal}
  {\bibinfo  {journal} {J. Low. Temp. Phys.}\ }\textbf {\bibinfo {volume}
  {67}},\ \bibinfo {pages} {115} (\bibinfo {year} {1987})}\BibitemShut
  {NoStop}%
\bibitem [{\citenamefont {Wilen}\ and\ \citenamefont
  {Giannetta}(1988)}]{Wilen1988}%
  \BibitemOpen
  \bibfield  {author} {\bibinfo {author} {\bibfnamefont {L.}~\bibnamefont
  {Wilen}}\ and\ \bibinfo {author} {\bibfnamefont {R.}~\bibnamefont
  {Giannetta}},\ }\href {\doibase 10.1007/BF00682147} {\bibfield  {journal}
  {\bibinfo  {journal} {J. Low. Temp. Phys.}\ }\textbf {\bibinfo {volume}
  {72}},\ \bibinfo {pages} {353} (\bibinfo {year} {1988})}\BibitemShut
  {NoStop}%
\bibitem [{\citenamefont {Shikin}(2002)}]{Shikin2002}%
  \BibitemOpen
  \bibfield  {author} {\bibinfo {author} {\bibfnamefont {V.}~\bibnamefont
  {Shikin}},\ }\href {\doibase 10.1134/1.1463111} {\bibfield  {journal}
  {\bibinfo  {journal} {JETP Letters}\ }\textbf {\bibinfo {volume} {75}},\
  \bibinfo {pages} {29} (\bibinfo {year} {2002})}\BibitemShut {NoStop}%
\bibitem [{sup()}]{supplemental}%
  \BibitemOpen
  \href@noop {} {\emph {\bibinfo {title} {See Supplemental Material at [URL
  will be inserted by publisher] for the details of our classical
  model}}}\BibitemShut {NoStop}%
\bibitem [{\citenamefont {Dykman}\ \emph {et~al.}(1997)\citenamefont {Dykman},
  \citenamefont {Fang-Yen},\ and\ \citenamefont {Lea}}]{Dykman1997}%
  \BibitemOpen
  \bibfield  {author} {\bibinfo {author} {\bibfnamefont {M.~I.}\ \bibnamefont
  {Dykman}}, \bibinfo {author} {\bibfnamefont {C.}~\bibnamefont {Fang-Yen}}, \
  and\ \bibinfo {author} {\bibfnamefont {M.~J.}\ \bibnamefont {Lea}},\ }\href
  {\doibase S0163-1829~97!05924-9} {\bibfield  {journal} {\bibinfo  {journal}
  {Phys. Rev. B}\ }\textbf {\bibinfo {volume} {55}},\ \bibinfo {pages} {16249}
  (\bibinfo {year} {1997})}\BibitemShut {NoStop}%
\bibitem [{\citenamefont {Teske}\ \emph {et~al.}(1999)\citenamefont {Teske},
  \citenamefont {Monarkha}, \citenamefont {Seck},\ and\ \citenamefont
  {Wyder}}]{Teske1999}%
  \BibitemOpen
  \bibfield  {author} {\bibinfo {author} {\bibfnamefont {E.}~\bibnamefont
  {Teske}}, \bibinfo {author} {\bibfnamefont {Y.~P.}\ \bibnamefont {Monarkha}},
  \bibinfo {author} {\bibfnamefont {M.}~\bibnamefont {Seck}}, \ and\ \bibinfo
  {author} {\bibfnamefont {P.}~\bibnamefont {Wyder}},\ }\href {\doibase
  S0031-9007(99)08755-4} {\bibfield  {journal} {\bibinfo  {journal} {Phys. Rev.
  Lett.}\ }\textbf {\bibinfo {volume} {82}},\ \bibinfo {pages} {2772} (\bibinfo
  {year} {1999})}\BibitemShut {NoStop}%
\bibitem [{\citenamefont {Monarkha}\ \emph {et~al.}(2000)\citenamefont
  {Monarkha}, \citenamefont {Teske},\ and\ \citenamefont
  {Wyder}}]{Monarkha2000}%
  \BibitemOpen
  \bibfield  {author} {\bibinfo {author} {\bibfnamefont {Y.~P.}\ \bibnamefont
  {Monarkha}}, \bibinfo {author} {\bibfnamefont {E.}~\bibnamefont {Teske}}, \
  and\ \bibinfo {author} {\bibfnamefont {P.}~\bibnamefont {Wyder}},\ }\href
  {\doibase ????} {\bibfield  {journal} {\bibinfo  {journal} {Phys. Rev. B}\
  }\textbf {\bibinfo {volume} {62}},\ \bibinfo {pages} {2593} (\bibinfo {year}
  {2000})}\BibitemShut {NoStop}%
\bibitem [{\citenamefont {Konstantinov}\ and\ \citenamefont
  {Kono}(2010)}]{Konstantinov2010}%
  \BibitemOpen
  \bibfield  {author} {\bibinfo {author} {\bibfnamefont {D.}~\bibnamefont
  {Konstantinov}}\ and\ \bibinfo {author} {\bibfnamefont {K.}~\bibnamefont
  {Kono}},\ }\href {\doibase 10.1103/PhysRevLett.105.226801} {\bibfield
  {journal} {\bibinfo  {journal} {Phys. Rev. Lett.}\ }\textbf {\bibinfo
  {volume} {105}},\ \bibinfo {pages} {226801} (\bibinfo {year}
  {2010})}\BibitemShut {NoStop}%
\bibitem [{\citenamefont {Chepelianskii}\ \emph {et~al.}(2015)\citenamefont
  {Chepelianskii}, \citenamefont {Watanabe}, \citenamefont {Nasyedkin},
  \citenamefont {Kono},\ and\ \citenamefont
  {Konstantinov}}]{Chepelianskii2015}%
  \BibitemOpen
  \bibfield  {author} {\bibinfo {author} {\bibfnamefont {A.~D.}\ \bibnamefont
  {Chepelianskii}}, \bibinfo {author} {\bibfnamefont {M.}~\bibnamefont
  {Watanabe}}, \bibinfo {author} {\bibfnamefont {K.}~\bibnamefont {Nasyedkin}},
  \bibinfo {author} {\bibfnamefont {K.}~\bibnamefont {Kono}}, \ and\ \bibinfo
  {author} {\bibfnamefont {D.}~\bibnamefont {Konstantinov}},\ }\href {\doibase
  10.1038/ncomms8210} {\bibfield  {journal} {\bibinfo  {journal} {Nature
  Communications}\ }\textbf {\bibinfo {volume} {6}},\ \bibinfo {pages} {7210}
  (\bibinfo {year} {2015})}\BibitemShut {NoStop}%
\end{thebibliography}
\end{document}